\documentclass[preprint2]{aastex62}

\graphicspath{{./}{figures/}}
\submitjournal{MNRAS}
\shorttitle{Detection of a Double Relic in SPT-CL J0245-5302}
\shortauthors{Zheng et al.}

\begin{document}

\title{Detection of a Double Relic in the Torpedo Cluster: SPT-CL J0245-5302}

\correspondingauthor{Qian Zheng}
\email{zq@bao.ac.cn}

\author[0000-0002-0786-7307]{Qian Zheng}
\affiliation{School of Engineering and Computer Science\\
  PO Box 600, Victoria University of Wellington\\
  Wellington 6140, New Zealand}
\affiliation{Shanghai Astronomical Observatory, Chinese Academy of Sciences\\ 
  80 Nandan Road, Shanghai 200030, China}

\author{Melanie Johnston-Hollitt}
\affiliation{Peripety Scientific Ltd.\\
 PO Box 11355, Manners Street, Wellington 6142, New Zealand}
\affiliation{International Centre for Radio Astronomy Research,Curtin University\\
Bentley, WA 6102, Australia}

\author{Stefan W. Duchesne}
\affiliation{Peripety Scientific Ltd.\\
 PO Box 11355, Manners Street, Wellington 6142, New Zealand}
\affiliation{International Centre for Radio Astronomy Research,Curtin University\\
Bentley, WA 6102, Australia}

\author{Weitian Li}
\affiliation{School of Physics and Astronomy, Shanghai Jiao Tong University\\
800 Dongchuan Road, Shanghai 200240, China}

\begin{abstract}
The Torpedo cluster, SPT-CL J0245-5302 (S0295) is a massive, merging cluster at a redshift of $z = 0.300$, which exhibits a strikingly similar morphology to the Bullet cluster 1E 0657-55.8 ($z = 0.296$), including a classic bow shock in the cluster's intra-cluster medium revealed by Chandra X-ray observations. We present Australia Telescope Compact Array data centred at 2.1 GHz and Murchison Widefield Array data at frequencies between 72 MHz and 231 MHz which we use to study the properties of the cluster. We characterise a number of discrete and diffuse radio sources in the cluster, including the detection of two previously unknown radio relics on the cluster periphery. The average spectral index of the diffuse emission between 70 MHz and 3.1 GHz is $\alpha=-1.63_{-0.10}^{+0.10}$ and a radio-derived Mach number for the shock in the west of the cluster is calculated as M = 2.04. The Torpedo cluster is thus a double relic system at moderate redshift.
\end{abstract}





\section{Introduction}
\label{sec:introduction}

Galaxy clusters are the largest known gravitationally bound structures in the Universe. Collisions between galaxy clusters provide an opportunity to understand the growth of structure in the Universe, constrain the nature and the properties of dark matter, and may be used as probes of cosmology. Multi-wavelength studies of clusters over the last 40 years have revealed many complex interacting systems which exhibit signs of mergers across the electromagnetic spectrum including elongated optical galaxy distributions \citep{Johnston-Hollitt08,Shakouri16}, morphologically disturbed X-ray emitting gas with evidence for shocks, cold fronts and temperature jumps \citep{Vikhlinin01, Markevitch01,Finoguenov06}, and spectacular diffuse radio emission $-$ the so-called `relics' and `haloes' (see \citealp{BrunettiJones14} for a review). Studying these systems has provided a reasonably coherent picture of cluster mergers which suggests the propagation of shock waves generated in mergers can be traced via detailed X-ray and radio imaging. Possibly the most famous and defining example of shock-driven emission in a galaxy cluster is 1E 0657-55.8 $-$ the Bullet cluster \citep{Markevitch02}. 

The Bullet cluster is a well known massive merging galaxy cluster, which is the result of two colliding galaxy clusters at a redshift of $z = 0.296$. The system contains a spectacular X-ray shock cone and associated temperature features \citep{Markevitch04} along with a radio halo \citep{Liang2000, Shimwell14}, and two peripheral radio relics, one of which appears to be the result of re-acceleration of the lobes of a radio galaxy \citep{Shimwell15}. Additionally, gravitational lensing studies of the Bullet cluster provide the best evidence for the existence of dark matter \citep{Clowe04,Markevitch04}. While many other cluster systems have been found to have elongated X-ray emission and double radio relics (e.~g. \citealp{mjh03, Bagchi06,vanWeeren11,Riseley17}), few have had the imaging fidelity of the Bullet cluster to reveal the precise shock-cone in the central plasma. 

Here we present a study of SPT-CL J0245-5302 (Abell S0295; \citealp{ACO89}) which is a merging galaxy cluster at $z = 0.300$ \citep{Ruel14}. SPT-CL J0245-5302 was initially detected via the Sunyaev-Zel'dovich effect (SZ) with a number of instruments including the Atacama Cosmology Telescope (ACT; \citealp{Fowler07}) and the South Pole Telescope (SPT; \citealp{Carlstrom11}). In this work, we use data from the Australia Telescope Compact Array (ATCA; \citealp{Frater92}) covering 1.1 to 3.1 GHz and the Murchison Widefield Array (MWA; \citealp{Tingay13}) at frequencies between 72 MHz and 231 MHz to study the properties of the cluster across the radio bands. We supplement this with Chandra X-ray imaging to study signatures of merging in the intra-cluster medium (ICM). We find evidence for large-scale diffuse radio emission as well as shock signatures in the X-ray imaging. The X-ray emission demonstrates that SPT-CL J0245-5302 has the same spectacular shock-cone evident in the X-ray imaging as the Bullet cluster. We thus dub SPT-CL J0245-5302 as the `Torpedo' cluster as the X-ray emission is reminiscent of a torpedo travelling at a speed greater than that of sound in the local medium.

This paper is laid out as follows: Section~\ref{sec:cluster} presents an overview of the cluster properties while Section~\ref{sec:observation} describes the observations and data reduction undertaken here. In Section~\ref{sec:results} we present a spectral analysis across the radio band, and we also study the properties of the diffuse radio emission detected. Finally, in Section~\ref{sec:conclusion} we summarise the results and make conclusions.

We adopt a standard set of cosmological parameters throughout with $H_0$ = $73 ~\mathrm{kms}^{-1}\mathrm{Mpc}^{-1}$, $H_0= 100 ~\mathrm{h}$, $\Omega_m$ = $0.27$, ${\Omega}_{\Lambda}$ = $0.73$. At the redshift of SPT-CL J0245-5302, 1$\arcsec$ corresponds to $\sim$2.95 $\mathrm{h}^{-1}$ kpc.

\section{The Torpedo Cluster}
\label{sec:cluster}
As mentioned above, the Torpedo cluster is first reported in the ACO cluster catalogue via detection as an optical over density \citep{ACO89} and the redshift of the cluster was first measured by \cite{Edge94} as 0.3006. This was confirmed by \cite{Ruel14} who determine an average redshift of $0.3001 \pm 0.0010$ and a velocity dispersion of $1235 \pm 211$ km s$^{-1}$ from 30 spectroscopic measurements. 
 
It was first detected in the X-ray in the ROSAT All-Sky Survey \citep{Voges99} and the Advanced Satellite for Cosmology and Astrophysics (ASCA) presented an average temperature of kT = 6.72 $\pm$ 1.09 keV \citep{Fukazawa04}. An X-ray luminosity of L$_x = 7.2 \pm 0.9 \times 10^{44}$ erg s$^{-1}$ was reported by \cite{Williamson11} in the soft X-ray band [0.5-2.0 keV]. An SZ detection was first presented in \cite{Hincks10} using ACT which was followed by confirmation with the SPT \citep{Williamson11, Bleem15, Lindner15}, with all SZ images showing an elongated structure along a northwest-southeast axis. Although the cluster was detected by the SPT at high significance, it was not included in the official catalogue due to its proximity to a strong radio point source. Nevertheless \cite{Williamson11} include it in their paper reporting M$_{500} \pm$ stat $\pm$ syst of $[8.1 \pm 2.1 \pm 1.5] \times 10^{14}$ M$_\odot$ h$^{-1}_{70}$ making it a relatively massive system.  

\cite{Menanteau10} performed detailed multi-coloured optical observations of the cluster using the Cerro Tololo Inter-American Observatory (CTIO) showing it to be extremely rich with a galaxy population which is highly elongated along the same northwest-southeast axis as the SZ imaging. Furthermore, they confirmed the presence of a large strong-lensing arc in the region of the brightest cluster galaxy that was first reported in \cite{Edge94}. These properties allow this to be unambiguously classified as a massive, merging system.

\section{Observations and data reduction}
\label{sec:observation}

\subsection{MWA Imaging}
The MWA is a low frequency radio interferometer located at the Murchison Radio-astronomy Observatory in Western Australia which is a protected radio-quiet site with very low levels of radio frequency interference (RFI) \citep{Offringa15}. At the time this work was undertaken, the Phase I MWA consisted of 128 tiles with 16 dipole antennas in each tile distributed over a 3 kilometre diameter area. The telescope operates between 72 MHz and 300 MHz, with a bandwidth of 30.72 MHz and further details of the instrument and the associated science may be found in \citet{Tingay13} and \citet{Bowman13}, respectively. One of the principal data products of the Phase I MWA is the GaLactic and Extragalactic All-sky MWA survey (GLEAM; \citealp{Wayth15}) which is a survey of the Southern sky ($\delta$  $\leq$ 30 degrees) at frequencies between 72 MHz and 231 MHz. 

In this work we used data from the GLEAM survey for the Torpedo cluster imaged at four frequencies: 87.5, 118.5, 154.5 and 200.5 MHz. The observations are described in \cite{Wayth15} and the data processing strategies follow those detailed in \cite{HurleyWalker14}.  A ``robust'' or ``Briggs'' \citep{briggs95} weighting scheme using a robustness parameter of $0.0$ was used to make the images and the final image properties including beam size and root mean square (rms) noise are presented in Table \ref{Tab:MWAimages}.


\begin{table}[h!]
\renewcommand{\thetable}{\arabic{table}}
\centering
\caption{Properties of the MWA images. $\nu_c$ is the centre frequency. The bandwidth of all observations is 30.72 MHz.} \label{Tab:MWAimages}
\begin{tabular}{ccc}
\tablewidth{0pt}
\hline
\hline
\decimals
 $\nu_c$ & RMS noise & FWHM beam\\ 
 (MHz) & (mJy/beam) & (arcmin)\\
\hline
87.5  & 24 &$5.07 \times 4.92$ \\
118.5  & 22 &$3.66 \times 3.58$ \\
154.5  & 13 &$2.80 \times 2.70$ \\
200.5  & 16 &$2.35 \times 2.22$ \\
\hline
\end{tabular}
\end{table}

\subsection{ATCA Imaging} 
\label{sec:atca}
The ATCA is an radio array of six 22 m antennas, which is located in the north-west of New South Wales. Observations of the Torpedo cluster (Project: C2837, PI: M. Johnston-Hollitt), were carried out in three different array configurations: EW352, 6D, and 1.5C, using a frequency bandwidth of 2048 MHz centred at 2.1GHz. Observations commenced in June 2013 and concluded in March 2017. We supplemented our data with archival observations in the 6A configuration collected in 2011 (Project: C2457, PI: A. Baker) to provide high resolution ($\sim 5\arcsec$) images to identify point sources. Details of the ATCA observations are presented in Table \ref{Tab:observations}.

\begin{deluxetable*}{ccccccc}
\tablecaption{Properties of the ATCA observations. t$_{\rm int}$ is the total integrated time on the cluster, $L_{max}$ is the longest baseline in each configuration with (without) antenna 6 and $\rm uv_{\rm max}$ is the corresponding length in kilowavelengths, similarly $L_{min}$ is the shortest baseline in each configuration, and $\rm uv_{\rm min}$ is the corresponding length in kilo wavelengths. The bandwidth is 2048 MHz and the central frequency is 2100 MHz.\label{Tab:observations}}
\tablewidth{0pt}
\tablehead{
  \colhead{Config.} &\colhead{Date} &\colhead{$t_{\rm int}$} &\colhead{$L_{\rm max}$} &\colhead{$\rm uv_{\rm max}$} &\colhead{$L_{\rm min}$} &\colhead{$\rm uv_{\rm min}$} \\
  \colhead{} &\colhead{} &\colhead{(mins)} &\colhead{(m)} &\colhead{k$\lambda$} &\colhead{(m)} &\colhead{k$\lambda$} 
}
\decimalcolnumbers
\startdata
EW352 &21,22-06-2013 & $210.7$  & 4439 (352) & 42.4 (3.1)& 31 & 0.22\\
1.5C  &21-12-2016    & $485.4$  & 4500 (1485) & 43.0 (14.2)& 77 & 0.54\\
H214  &15-3-2017     & $871.9$  & 4500 (247) &43.0 (2.4)&  92 & 0.64\\ 
6A    &12-12-2011    & $625.8$  & 5939 (2923) &56.8 (27.9)& 337 & 2.36\\
\enddata
\end{deluxetable*}


The data reduction followed standard procedures for CABB continuum data reduction with {\sc miriad} \citep{Sault95}. Briefly, the following procedures were undertaken: first data were flagged to remove RFI and the edge channels that are within the bandpass roll-off, which accounts for about 32 MHz on each end of the band. Each dataset was then split into four sub-bands of $\sim$500 MHz width to improve calibration and deconvolution in a procedure commonly used for wide-band ATCA data reduction with {\sc miriad}  (e.~g. \citealp{Shakouri16b, Martinez16, Martinez17, Duchesne17}). Next, the flux density scale was set relative to the unresolved calibrator PKS B1934-638 which was also used for the bandpass correction. The flux density of PKS B1934-638 at 1.4 GHz is 14.95 Jy as detailed in the Australia Telescope National Facility documentation. The unresolved source PKS B0252-549 was used as the phase calibrator in our observations (Proj. C2837) while PKS B0214-522 was used for the archival 6A observations (Proj. C2457). 

\subsection{Low Resolution Imaging}
\begin{figure}[ht!]
\begin{center}
\includegraphics[angle=0,scale=0.1]{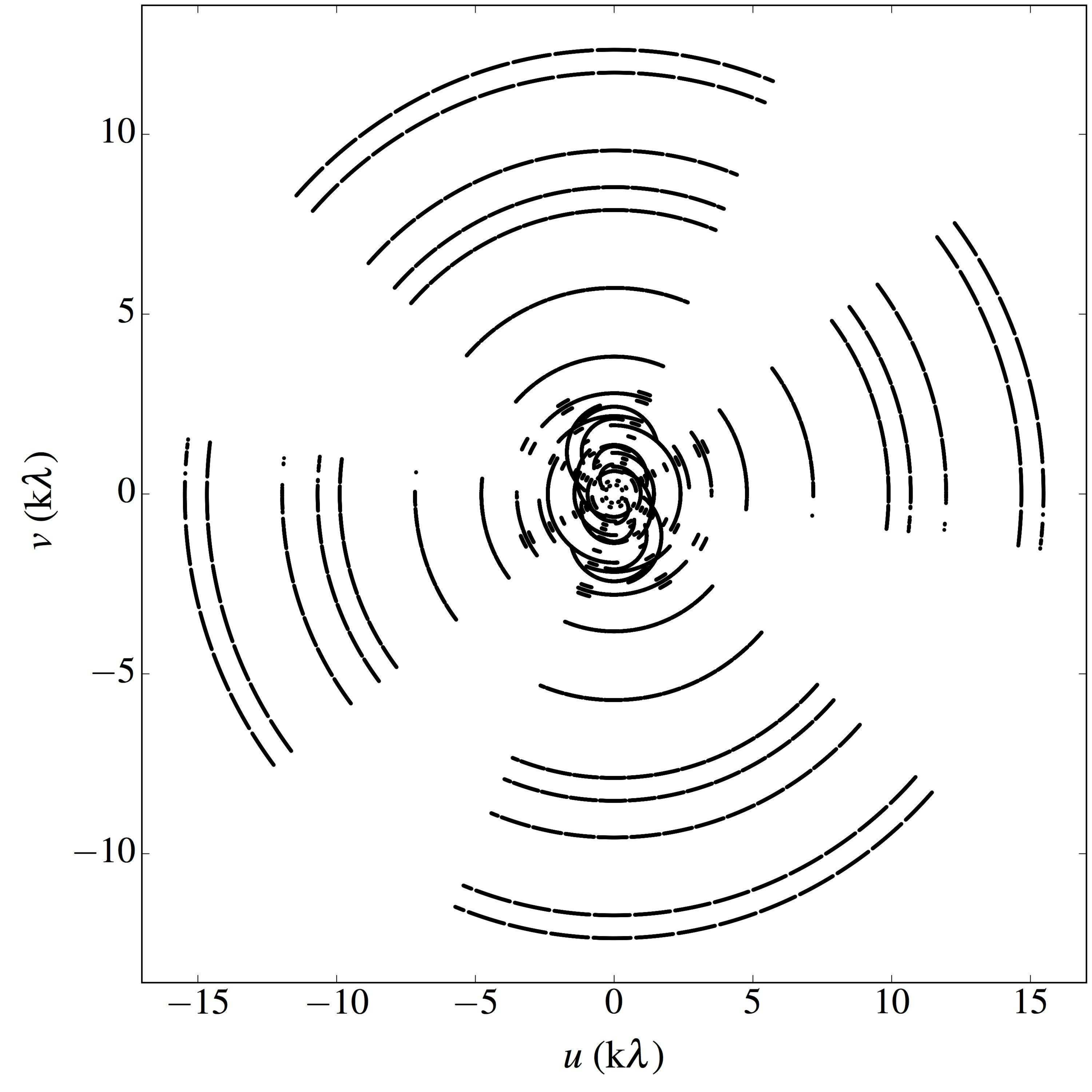} 
\end{center}
\caption{The uv-coverage of the low resolution image at 2868 MHz.\label{fig:uv-low}}
\end{figure}

As the diffuse components in clusters are observed at comparatively large physical and hence angular scales, we used the combined H214, EW352 and 1.5C configurations without the baselines to antenna 6 to make a set of low resolution images of each frequency sub-band. Final Stokes I images of the four sub-bands were centred at 1332, 1844, 2356, and 2868 MHz, and imaging was performed with a robust weighting scheme with a parameter of 1.0, so as to reduce the effect of side lobes. The calibrated data were imaged using the standard CLEAN algorithm, with a threshold of 3 sigma (three times of the root mean square noise). The procedure was as follows: multi-frequency CLEAN (mfclean), then one round of amplitude + phase self calibration with selfcal, and then another round of mfclean before restoring. The final low resolution image properties are given in Table \ref{Tab:images-low}. Note that the beam for the lowest frequency sub-band is considerably elongated compared to the other sub-bands, and the rms noise is considerably higher. This is the result of removal of large amounts of RFI in the lowest part of the ATCA 2.1 GHz band. This is a common problem with ATCA imaging often resulting in the the use of the second lowest frequency band in detection experiments, even for steep spectral index sources such as haloes and relics in galaxy clusters (e.~g. \citealp{Shakouri16b, Martinez16}). Figure \ref{fig:uv-low} presents the resultant uv-coverage for the low resolution images.

\begin{table}[h!]
\renewcommand{\thetable}{\arabic{table}}
\centering
\caption{Properties of the low resolution ATCA sub-band images made from combining the H214, EW352 and 1.5C configurations without the baselines to antenna 6. $\nu_c$ is the centre frequency and pa is the position angle of the synthesised beam (point spread function).}\label{Tab:images-low}
\begin{tabular}{cccc}
\tablewidth{0pt}
\hline
\hline
$\nu_c$ & RMS noise & FWHM beam & pa\\ 
 (MHz)   & (mJy/beam) & (arcsec)& (degrees)\\
\hline
1332 &0.35  &$25.4\times21.4$ &20.5\\  
1844 &0.14  &$23.2\times16.2$ &63.1\\   
2356 &0.11  &$15.5\times13.2$ &45.1\\
2868 &0.10  &$18.2\times10.6$ &70.2\\
\hline  
\end{tabular}
\end{table}

\subsection{High Resolution Imaging}
\begin{figure}[ht!]
\begin{center}
\includegraphics[angle=0,scale=0.1]{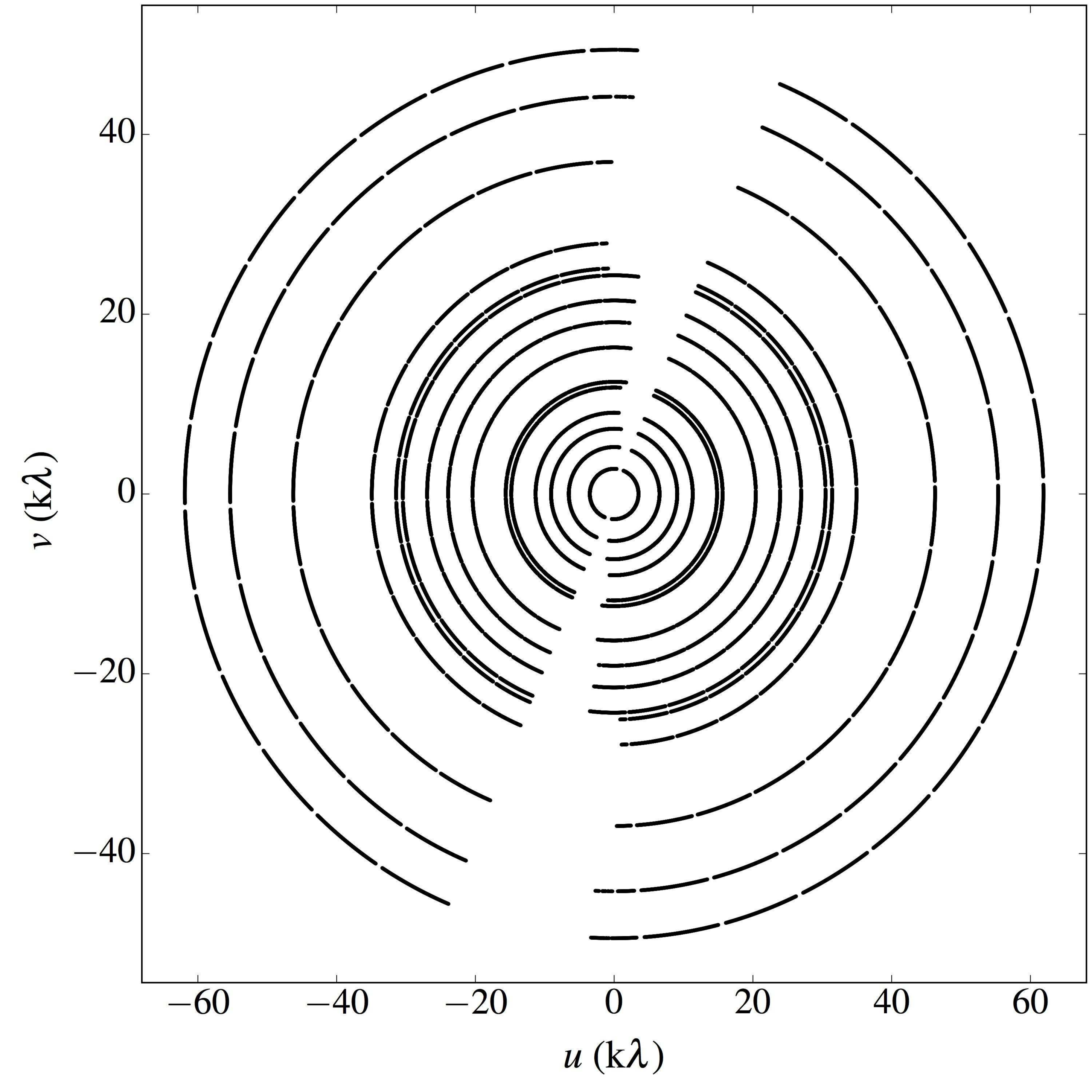} 
\end{center}  
\caption{The uv-coverage of the high resolution image at 2868 MHz.\label{fig:uv-high}}
\end{figure}

In order to separate diffuse low surface brightness components in the cluster from discreet compact sources we created a set of high resolution images using the archival 6A data. Following an identical imaging process, sub-band image were made at the same central frequencies as the low resolution imaging, but without application of a taper. The final high resolution image properties are given in Table \ref{Tab:images-high} and Figure \ref{fig:uv-high} presents the uv-coverage.

\begin{table}[h!]
\renewcommand{\thetable}{\arabic{table}}
\centering
\caption{Properties of the high resolution ATCA sub-band images using 6A configuration. $\nu_c$ is the centre frequency and pa is the position angle of the synthesised beam (point spread function).}\label{Tab:images-high}
\begin{tabular}{cccc}
\tablewidth{0pt}
\hline
\hline
$\nu_c$ & RMS noise & FWHM beam & pa\\ 
 (MHz)   & (uJy/beam) & (arcsec)& (degrees)\\
\hline
1332 & 40.4 &$7.4\times4.5$ &3.4\\  
1844 & 29.3 &$5.3\times3.4$ &5.4\\  
2356 & 27.6 &$4.3\times2.6$ &4.8\\
2868 & 29.1 &$3.6\times2.3$ &5.5\\
\hline  
\end{tabular}
\end{table}

\subsection{Chandra Imaging}
We used 5 archival observations towards our target cluster obtained by the Chandra X-ray Observatory with its Advanced CCD Imaging Spectrometer (ACIS). The observations are listed in Table \ref{Tab:chandra}. Version 4.9 of the CIAO software\footnote{\url{http://cxc.cfa.harvard.edu/ciao/}} with version 4.6.2 of the calibration database `CALDB' were used to perform the data reduction and analysis. 

For each observation, the raw data downloaded from the Chandra Data Archive were reduced with the "\texttt{chandra\_repro}" tool to create the calibrated level=2 event file, which was further cleaned using the "\texttt{lc\_clean}" tool to clip the times affected by flares due to solar activity. The cleaned total exposure time is $\sim 196.7$ ks, making this a high-quality data set suitable to reveal the fine structures within the cluster ICM.

\begin{figure*}[ht!]
\plotone{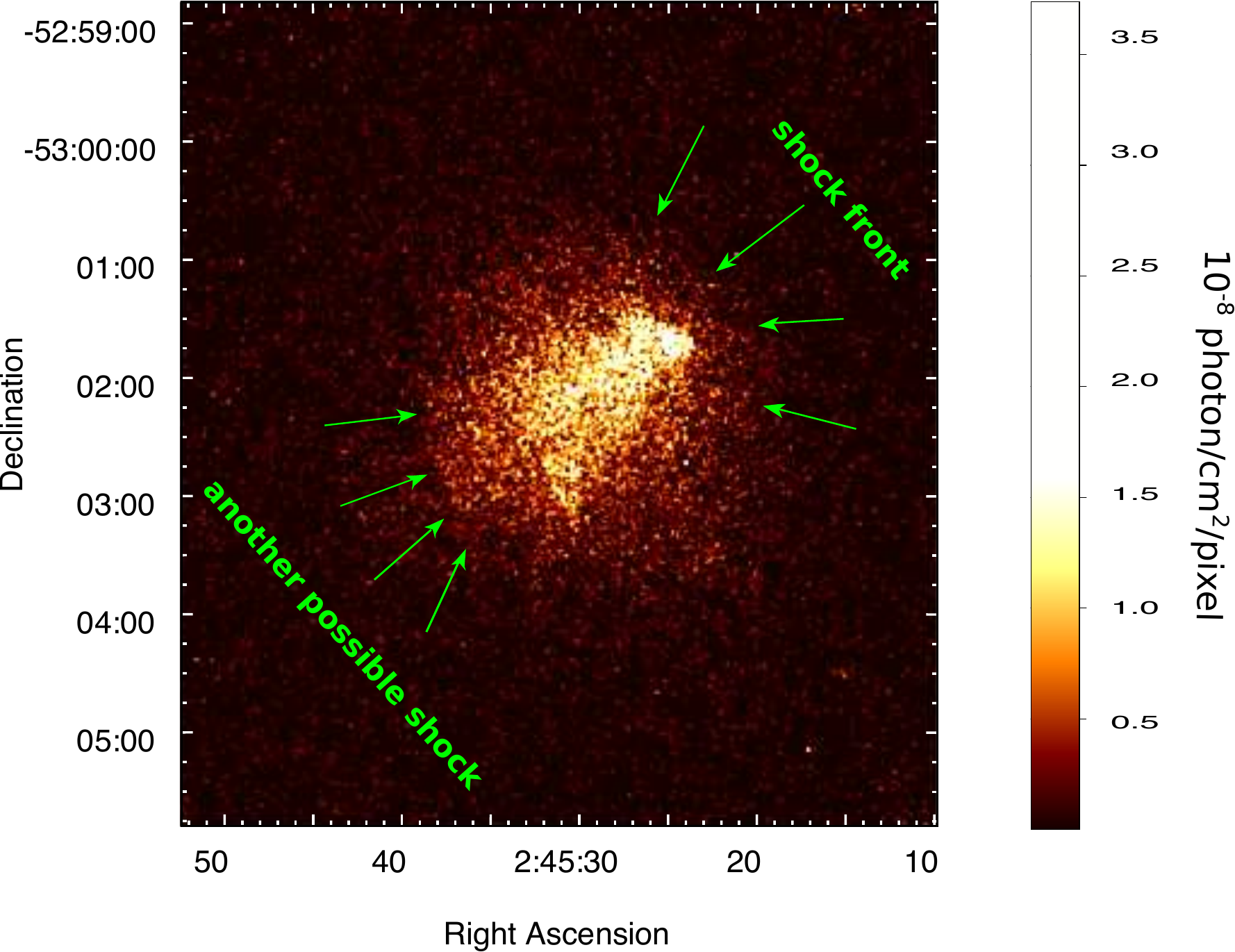}
\caption{The 0.5-2.0 keV soft X-ray image of SPT-CL J0245-5302 observed with the Chandra ACIS detector. The image was corrected for exposure and the point sources were removed. The units are photon/cm$^2$/s/pixel and the resolution of the Chandra image is $\sim$ 0.492 arcsec/pixel. The presence of a shock is visible in the northwest of the cluster, and is very similar to the morphology of the Bullet cluster. A second possible shock is also seen in the southeast of the cluster.\label{fig:softX}}
\end{figure*}

We then combined all the cleaned event files and made an exposure-corrected image within the soft X-ray band (0.5-2.0 keV) using the "\texttt{merge\_obs}" tool. The bright point sources in the image were detected with the "\texttt{celldetect}" tool, visually examined, removed, and refilled by sampling from their neighboring regions with the "\texttt{dmfilth}" tool. The final exposure-corrected and source-removed soft-band (0.5-2.0 keV) image of the Torpedo cluster, SPT-CL J0245-5302, is shown in Figure \ref{fig:softX}. The resolution of the final Chandra image is $\sim$ 0.492 arcsec/pixel. A break in the X-ray surface brightness can be found in the northwest of the cluster in Figure \ref{fig:softX}, and the break is caused by a projected abrupt gas density discontinuity. Another break can be seen in the southeast of the cluster.

\begin{table}[h!]
\renewcommand{\thetable}{\arabic{table}}
\centering
\caption{Five archived Chandra X-ray observations used in this work. See \url{http://cxc.cfa.harvard.edu/cda/}}\label{Tab:chandra}
\begin{tabular}{ccccc}
\tablewidth{0pt}
\hline
\hline
 No. & ObsID & Date & Instrument & Exposure\\ 
   & &  & & (ks)\\
\hline
1 &  12260 & 2012-Jan-06 & ACIS-I & 19.79 \\
2 &  16127 & 2014-Jul-25 & ACIS-I & 43.32 \\
3 &  16524 & 2014-May-20 & ACIS-I & 44.60 \\
4 &  16525 & 2014-May-17 & ACIS-I & 44.48 \\
5 &  16526 & 2014-Aug-13 & ACIS-I & 44.48 \\
\hline  
\end{tabular}
\end{table}

\section{Results}
\label{sec:results}

\begin{figure}[ht!]
 \begin{center}
\includegraphics[angle=0,scale=0.55]{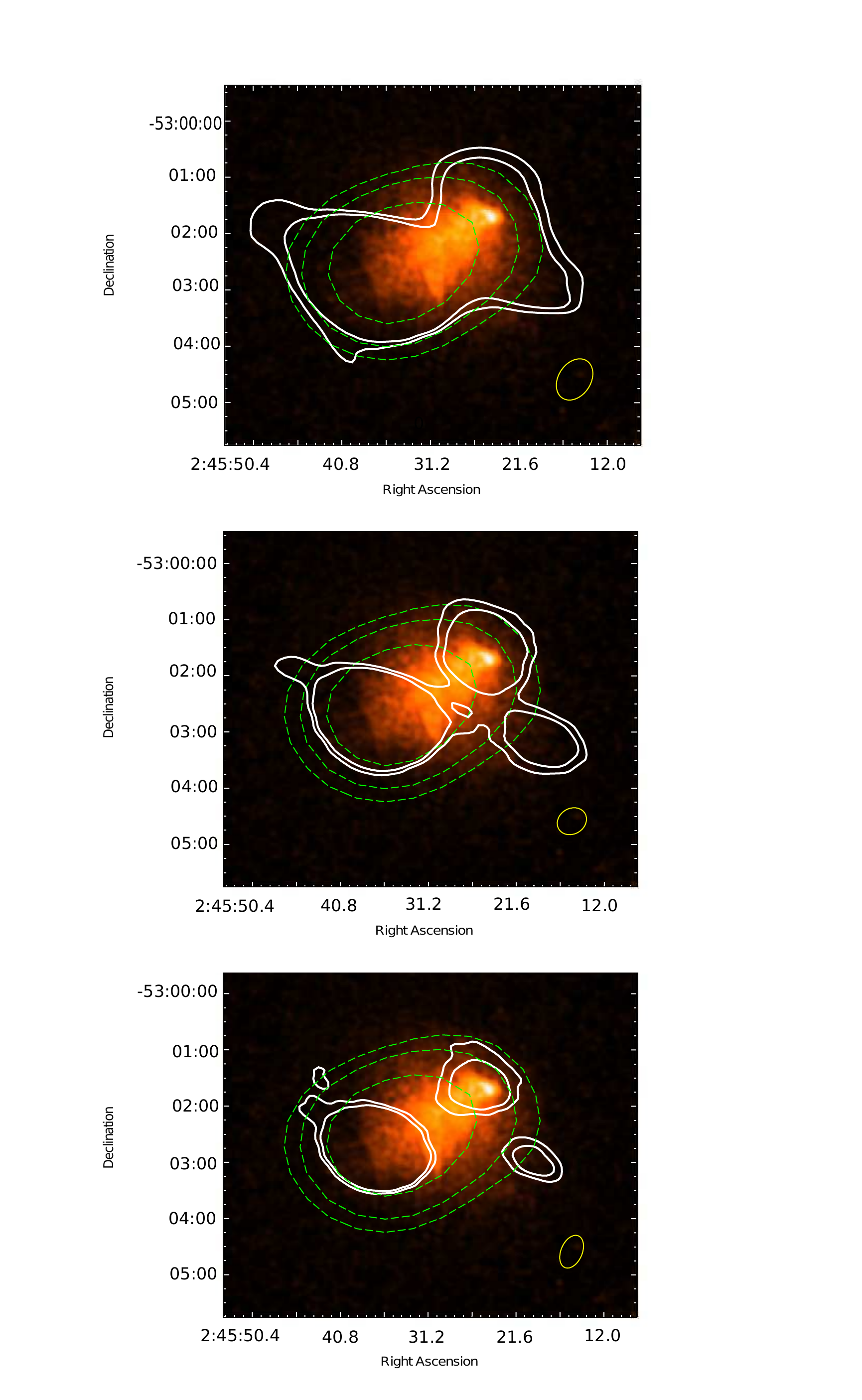} 
\end{center}
\caption{Chandra soft-band X-ray images of SPT CL J0245-5302 overlaid with the contours of the 200 MHz MWA image (green: 2$\sigma$, 3$\sigma$, and 5$\sigma$) and the contours of the low resolution ATCA images at 1844, 2356 and, 2868 MHz (white: 3$\sigma$, and 5$\sigma$) from top to bottom, respectively: The 1332 MHz frequency band is not presented due to the poor sensitivity and uv-coverage of the ATCA image at that frequency. Note the sharp shock cone evident in the X-ray emission. The yellow circles present in the right bottom of the panels show the synthesised beams of the ATCA images.\label{fig:ATCAimages}}
\end{figure}

Figure \ref{fig:ATCAimages} shows the Chandra soft-band X-ray image of SPT-CL J0245-5302 overlaid with the 200 MHz MWA contours (green) and contours from three of the four low resolution ATCA sub-band images (1844, 2356, and 2868 MHz). Due to the extensive RFI in the lowest frequency sub-band we do not show it here. The synthesised beams of the ATCA images are shown in the right bottom of each panel in Figure \ref{fig:ATCAimages}. The synthesised beams of the MWA image is too large to be presented, but is approximately $2.35\times2.22$ arcmin (200 MHz) as shown in Table \ref{Tab:MWAimages}.

The Chandra imaging shows the X-ray emission has a classic bow shock evident in the northwest of the cluster, identical to that seen in the Bullet cluster \citep{Markevitch01}. The X-ray gas follows both the optical galaxy distribution presented in \cite{Menanteau10} and the SZ elongation \citep{Hincks10,Williamson11, Bleem15, Lindner15}. The MWA finds strong radio emission coincident with the X-ray distribution and following the same elongated northwest-southeast axis. 

In the ATCA images, we find large scale emission across the cluster which resolves into discreet components as the resolution increases across the ATCA band. At the highest resolution (2668 MHz) we find three sources of radio emission. In the east of the cluster there is emission which is coincident with a known radio source SUMSS J024536$-$530244 \citep{Mauch03} (called Source 6 in Section \ref{sec:dis}) and to the west there are two bright patches of the emission, one to the northwest of the cluster coincident with the shock region shown in the X-ray image. The other emission is in the southwest of the cluster outside of the X-ray emitting region. At lower frequencies the two patches of emission on the western side of the cluster connect. In addition to the sources found here there is a known radio source in the cluster core: SUMSS J024525-530136 which was detected at 843 MHz \citep{Mauch03} (called Source 1 in Section \ref{sec:dis}). 

Comparison with the high resolution imaging (Figure \ref{fig:6A}) reveals that although there are a number of discrete radio sources in the cluster, much of this emission presented in the ATCA low resolution imaging is diffuse. In the following subsections we discuss how we separate the diffuse and discrete sources to allow both flux density estimation and imaging of the diffuse components.

\subsection{Identification of Discrete Point Sources}
\label{sec:dis}
\begin{figure*}[ht!]
\plotone{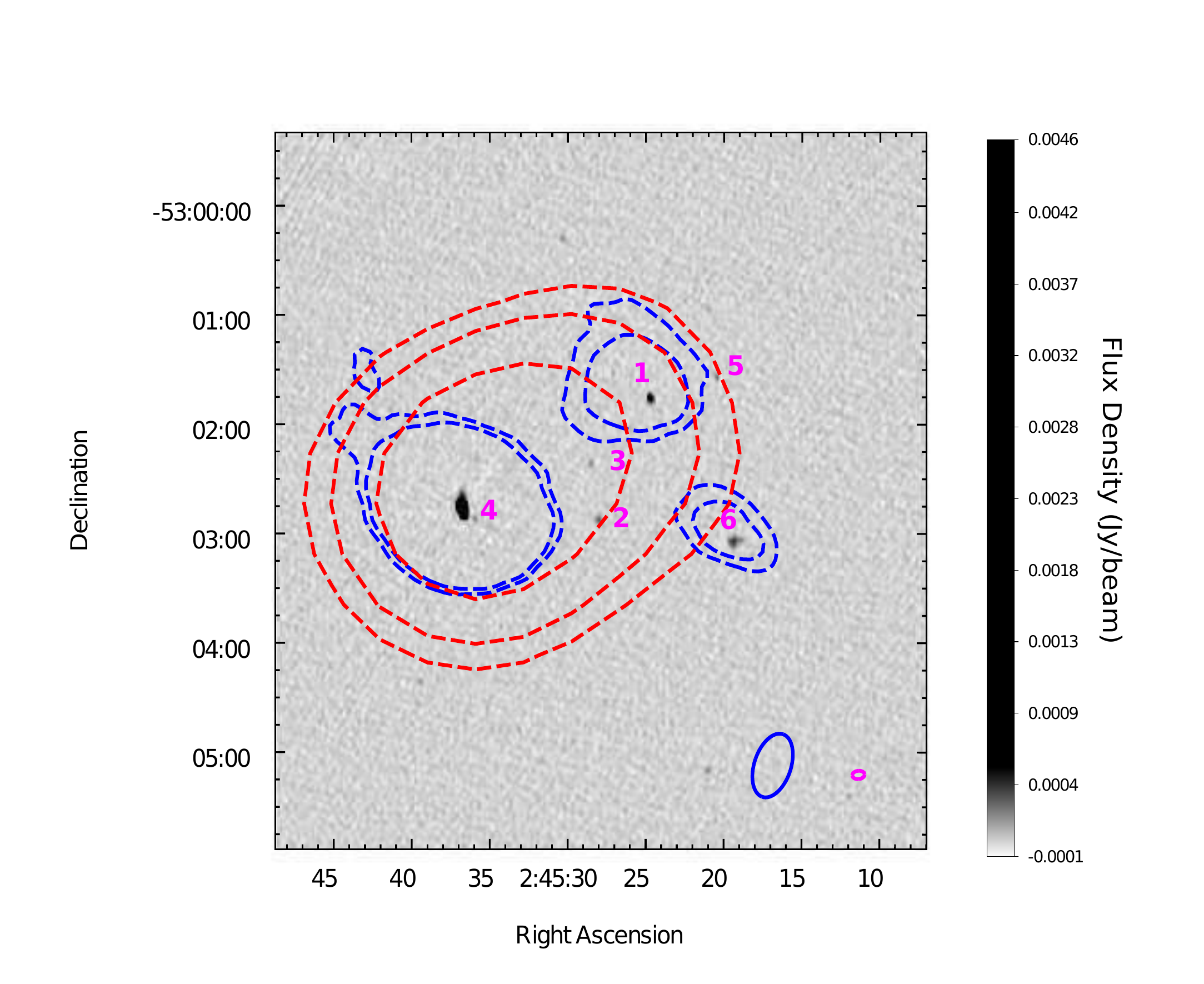}
\caption{High resolution 2868 MHz ATCA image of SPT CL J0245-5302 using the 6A configuration overlaid with the contours of the 200 MHz MWA image (red; shown at 2$\sigma$, 3$\sigma$, and 5$\sigma$) and the contours of the low resolution 2868 MHz ATCA image made using the EW352, 1.5C and H214 configurations without antenna 6 (blue; shown at 3$\sigma$ and 5$\sigma$). The high and low resolution ATCA synthesised beams are shown as the pink and blue ellipses in the right bottom of the figure, respectively. As before, the MWA synthesised beam is too large to display. The six discrete sources detected in the high resolution ATCA image are marked with pink numbers.\label{fig:6A}}
\end{figure*}

As stated above, there is diffuse emission evident in the low resolution ATCA images, but there are also contaminating discrete point sources. To distinguish the discrete point sources from the diffuse emission in Figure \ref{fig:ATCAimages} we used the high resolution ATCA images made using the 6A configuration. Figure \ref{fig:6A} shows the highest available resolution ATCA image (the 2868 MHz sub-band image made with the 6A configuration) overlaid with the MWA 200 MHz contours (red) and the corresponding low resolution 2868 MHz contours (blue) taken from the H214, EW352 and 1.5C configurations without data from antenna 6. 

\begin{figure*}[ht!]
\plotone{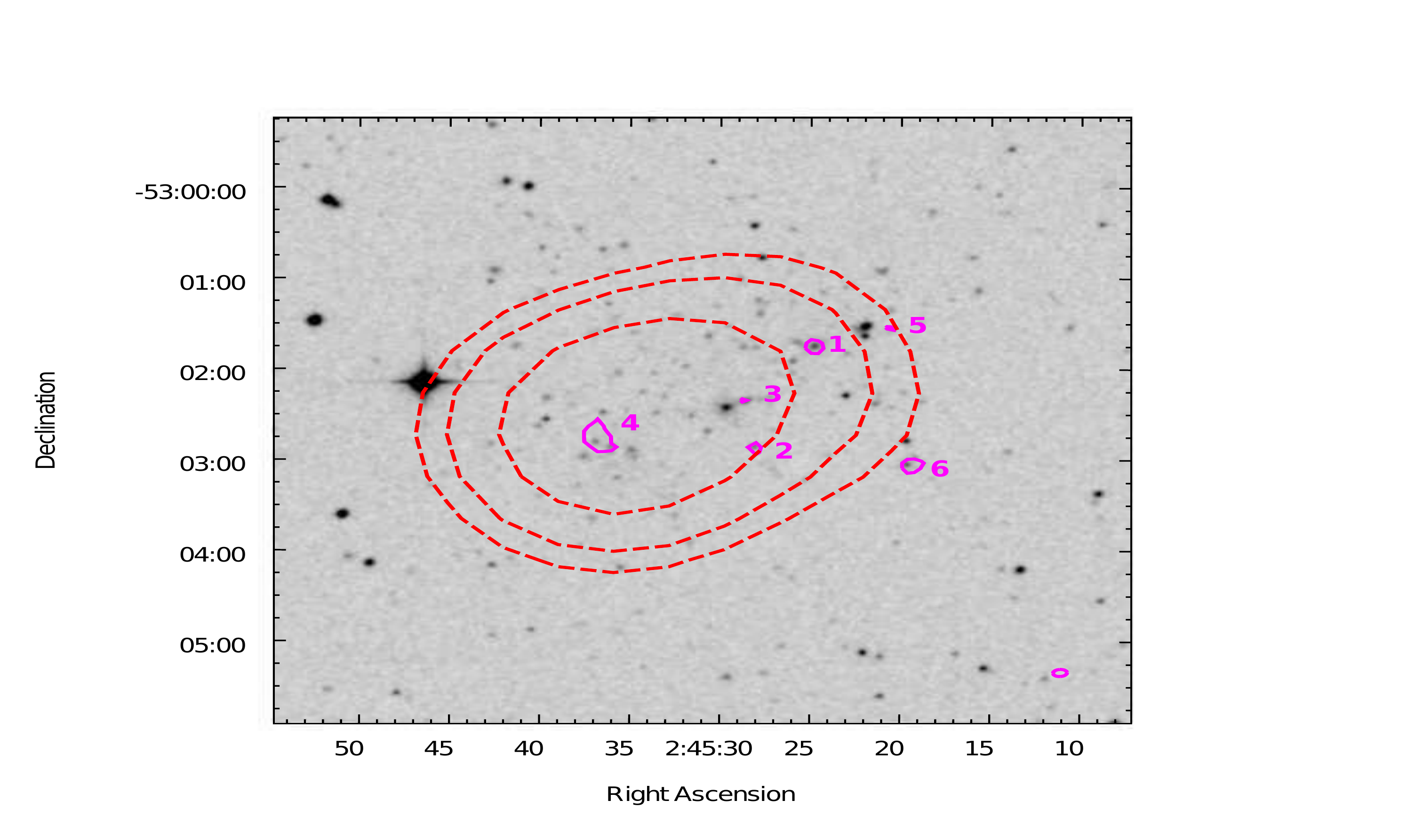}
\caption{The 2868 MHz high resolution contours (pink) of the six sources displayed in Figure \protect\ref{fig:6A} shown on an optical image of the Torpedo cluster from the DSS. The red contours are the 2$\sigma$, 3$\sigma$ and 5$\sigma$ contours of the 200 MHz MWA image. The synthesised beam of the ATCA images is shown at the bottom right.\label{fig:Opt-pointsource}}
\end{figure*}

\begin{figure}[ht!]
  \begin{center}
\includegraphics[angle=0,scale=0.4]{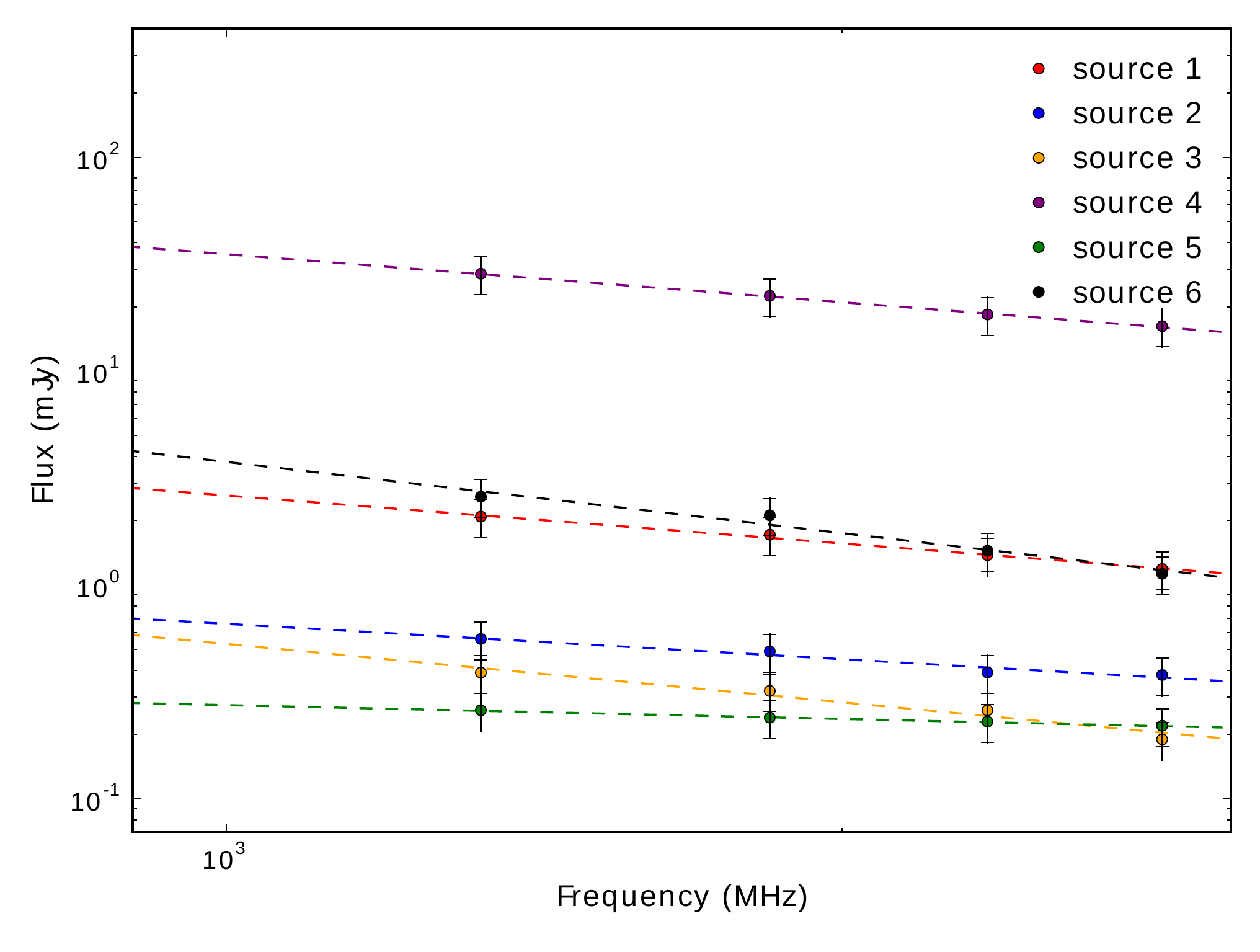} 
\end{center}
\caption{The fitted spectra of the six sources displayed in Table \ref{tab:sources}.\label{fig:pointsources}}
\end{figure}

Six discrete sources are identified in the cluster, which are shown in Figure \ref{fig:6A} and listed in Table \ref{tab:sources} as Sources 1, 2, 3, 4, 5 and 6. The six sources are also displayed on an optical image from the Digitized Sky Survey 2 (DSS2) in Figure \ref{fig:Opt-pointsource}. The measured flux densities of the sources at each frequency band taken from the high resolution images are also presented in Table \ref{tab:sources}. In addition we fit the radio spectral profile of the sources using the power-law relation $S\sim\nu^{\alpha}$ using errors of $20\%$ of the measured flux in the fitting. The fitting results of the spectral profiles of the sources are shown in Figure \ref{fig:pointsources} and the resultant spectral indices $\alpha$ are given in Table \ref{tab:sources}. 

\begin{figure}[ht!]
\begin{center}
\includegraphics[angle=0,scale=0.4]{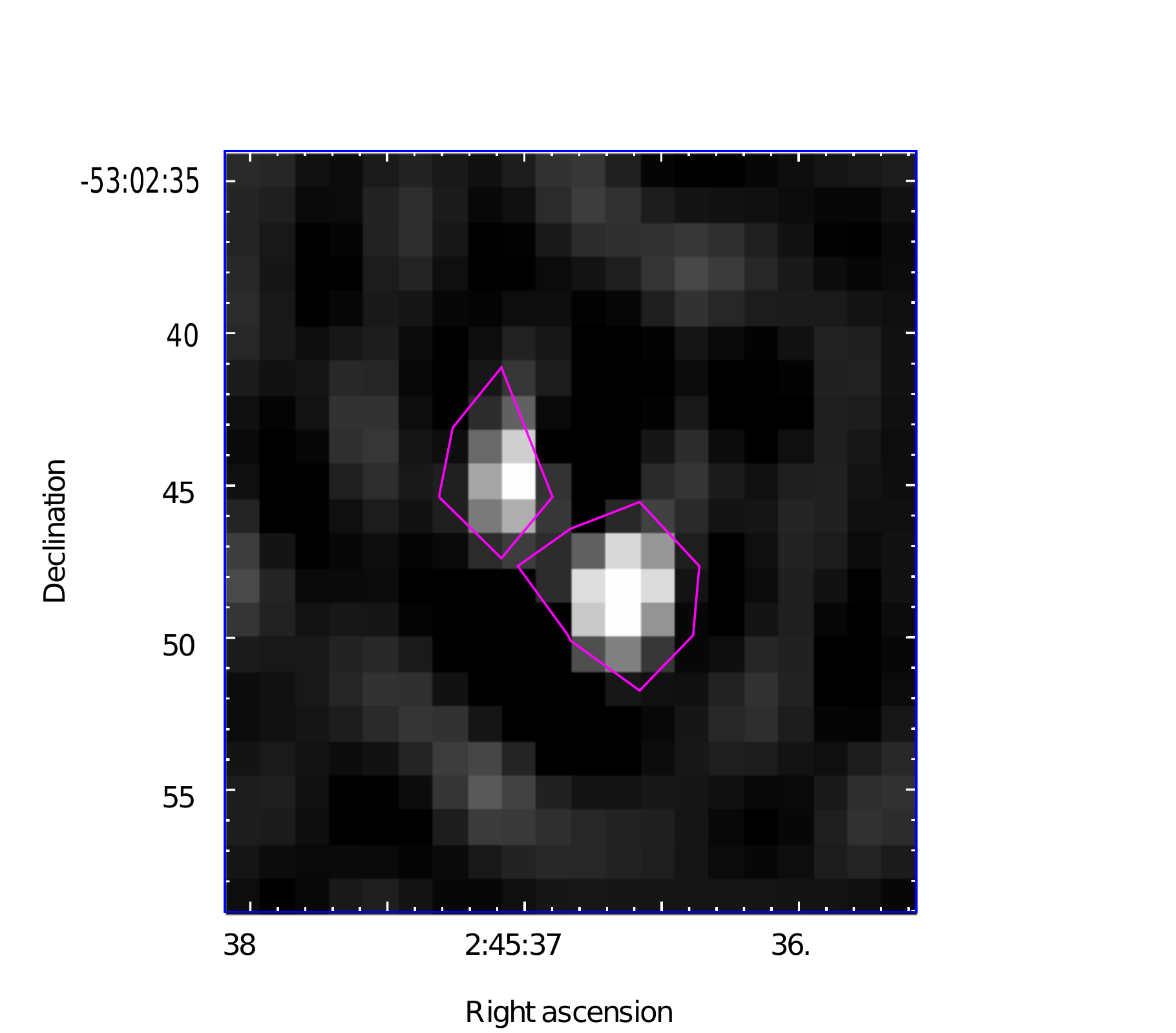} 
\end{center}
\caption{Source 4 imaged at 2868 MHz using the longest baselines in the 6A configuration data. The image has a resolution of 2.80 by 1.84 arc seconds and shows the source is comprised of two components.\label{source4}}
\end{figure}

We note that Source 4 has extended structure in the high resolution 2868 MHz image. Figure \ref{source4} displays Source 4 imaged using only the longest baseline in the 6A configuration providing a lower sensitivity but highest resolution image 2.803 by 1.841 arcseconds. This figure shows Source 4 to be further resolved into two components. Unfortunately, it is impossible to accurately measure the flux densities of these components separately in the other images due to the limitations of the resolution and uv-coverage. As a result the flux densities of Source 4 presented in Table \ref{tab:sources} are the sum of the two components and the spectral index is an average of these two components.

Source 6 has a very steep spectrum $\alpha$ = $-1.11_{-0.12}^{+0.06}$ and appears to be surrounded by diffuse emission as seen in the low resolution imaging. Conversely, Source 5 has a very flat spectral index across the ATCA band $\alpha$ = $-0.21_{-0.12}^{+0.05}$ and is extremely compact, suggestive of a compact flat-spectrum active galactic nuclei (AGN) \citep{Koay11}. The remaining sources have spectral indices in the range of -0.91 $\leq \alpha \leq$ -0.55 and are thus fairly typical AGN.

We also list the names of optical IDs for each source from the literature in Table \ref{tab:sources}. While all of the sources have optical counterparts, only five have been presented in the literature. Spectroscopically measured redshifts are available for four of the six radio sources (Source 1, 4, 5 and 6) and all fall within the redshift range 0.2966 $\leq z \leq$ 0.3033 \citep{Ruel14}, thus making them members of the Torpedo cluster which has an average redshift of z=0.300. Although the redshifts of Source 2 and 3 are unknown, examination of the photometric data from CTIO \citep{Menanteau10} suggests these sources are also cluster members. 

Most of the discrete sources in the Torpedo cluster are unresolved and very weak with flux densities at 1332 MHz of less than 2.6 mJy and are unlikely to contribute to the considerable emission detected in the low resolution imaging. Source 6, however, is both partially resolved and relatively bright with a flux density of 28.56 mJy at 1332 MHz. This source is the previously identified radio galaxy SUMSS J024536$-$530244, which is associated with the optical host galaxy MRSS 154-066225 (02h 45m 19.33s, -53d 03m 06.18s). This galaxy is in the cluster and has a measured spectroscopic redshift 0.296600. In the ATCA imaging it appears as a double-lobed source at the highest resolution (Figure \ref{source4}) and the low resolution images (Figure \ref{fig:ATCAimages}) suggest there is extended diffuse emission surrounding it. The spectral index is very steep $-1.11_{-0.12}^{+0.06}$, which is also consistent with the steep spectra typical of diffuse emission in clusters. 

In order to isolate the diffuse emission from these discrete sources we need to account for the flux density contributions of discrete sources at both MWA and ATCA frequencies and to understand the true morphology of the diffuse emission we need to remove the discrete sources from the visibility data and re-image the low resolution ATCA images. This will be discussed in the next two subsections. 

\begin{deluxetable*}{cccccccll}
\tablecaption{Details of the discrete radio sources detected in the Torpedo cluster. Column 1 is the source number as given on Figure \protect\ref{fig:Opt-pointsource}, column 2 is the radio-derived coordinates in J$_{2000}$, columns 3 and 4 are the optical host name and spectroscopic redshift from \protect\cite{Ruel14}, columns 5 through 8 are flux densities across the ATCA band and finally column 9 is fitted spectral indices $\alpha$ ($S\sim\nu^{\alpha}$, $\nu$ is the frequency). Sources 1, 2, 3, 4, 5 are point sources, Source 6 shows diffuse emission with a steep spectral index.\label{tab:sources}}
\tablewidth{700pt}
\tabletypesize{\scriptsize}
\tablehead{
  \colhead{ID} & \colhead{RA, DE} & \colhead{Optical ID} & \colhead{z} & \colhead{$\rm S_{1332}$} & \colhead{$\rm S_{1844}$} & \colhead{$\rm S_{2356}$} & \colhead{$\rm S_{2868}$} & \colhead{$\alpha$} \\
  \colhead{} & \colhead{(J2000)} & \colhead{} & \colhead{} & \colhead{(mJy)} & \colhead{(mJy)} & \colhead{(mJy)} & \colhead{(mJy)} & \colhead{}
}
\decimalcolnumbers
\startdata
1 &02h 45m 24.83s, -53d 01m 45.25s& APMUKS(BJ) B024349.99-531419.4 & 0.3028 &2.09$_{-0.42}^{+0.42}$& 1.72$_{-0.34}^{+0.34}$ & 1.38$_{-0.28}^{+0.28}$ &  1.19$_{-0.24}^{+0.24}$ & $-0.74_{-0.09}^{+0.09}$\\
\\
2 &02h 45m 28.07s, -53d 02m 52.08s& MRSS 154-065683&- &0.56$_{-0.11}^{+0.11}$ & 0.49$_{-0.10}^{+0.10}$ &  0.39$_{-0.08}^{+0.08}$ &  0.38$_{-0.08}^{+0.08}$ & $-0.55_{-0.09}^{+0.06}$\\
\\
3 &02h 45m 28.64s, -53d 02m 21.08s& - &- &0.39$_{-0.08}^{+0.08}$ & 0.32$_{-0.06}^{+0.06}$ &  0.26$_{-0.05}^{+0.05}$ &  0.19$_{-0.04}^{+0.04}$ & $-0.91_{-0.09}^{+0.06}$\\
\\
4 &02h 45m 36.64s, -53d 02m 48.23s& J024536.91-530248.1& 0.3021 &28.56$_{-5.71}^{+5.71}$ &22.49$_{-4.50}^{+4.50}$ & 18.40$_{-3.68}^{+3.68}$ & 16.24$_{-3.25}^{+3.25}$ & $-0.75_{-0.11}^{+0.06}$\\
\\
5 &02h 45m 20.53s, -53d 01m 33.65s& [RBB2014] J024518.99-530129.8& 0.3033&0.26$_{-0.05}^{+0.05}$ & 0.24$_{-0.05}^{+0.05}$ &  0.23$_{-0.05}^{+0.05}$ &  0.22$_{-0.04}^{+0.04}$ & $-0.21_{-0.12}^{+0.05}$\\
\\
6 &02h 45m 19.33s, -53d 03m 06.18s& MRSS 154-066225 & 0.2966 &2.59$_{-0.52}^{+0.53}$ & 2.12$_{-0.42}^{+0.42}$ &  1.45$_{-0.29}^{+0.29}$ &  1.13$_{-0.23}^{+0.23}$ & $-1.11_{-0.12}^{+0.06}$\\
\enddata
\end{deluxetable*}

\subsection{Discrete Source Subtraction in Spectra}
In both the MWA and low resolution ATCA images we detect significant emission which is a combination of the 6 discrete sources in the cluster and the diffuse emission detected in the low resolution ATCA imaging. In order to account for the contribution of the discrete sources to the total measured flux densities in both the MWA and low resolution ATCA sub-band images, we need to subtract the flux densities of the discrete sources. In the case of the ATCA data this could be accomplished by directly subtracting the flux densities measured in the high resolution sub-band images from the total flux density measured in the low resolution images. For the MWA data, which is at considerably low frequencies, we integrated the whole flux density within the envelope of the diffuse component in the MWA images. We then extrapolated the fitted power-law of the discrete sources to the MWA frequency bands and subtracted the flux densities of the discrete sources within the integration area to get the integrated flux density of the diffuse component. As the flux scale is not perfectly constrained in the MWA images and we have incomplete uv-coverage in the ATCA images, we consider $20\%$ flux errors for the flux estimated from the images, which is the error of the flux measured in the MWA images used in \cite{Hindson14}.

\begin{figure}[ht!]
\begin{center}
\includegraphics[angle=0,scale=0.4]{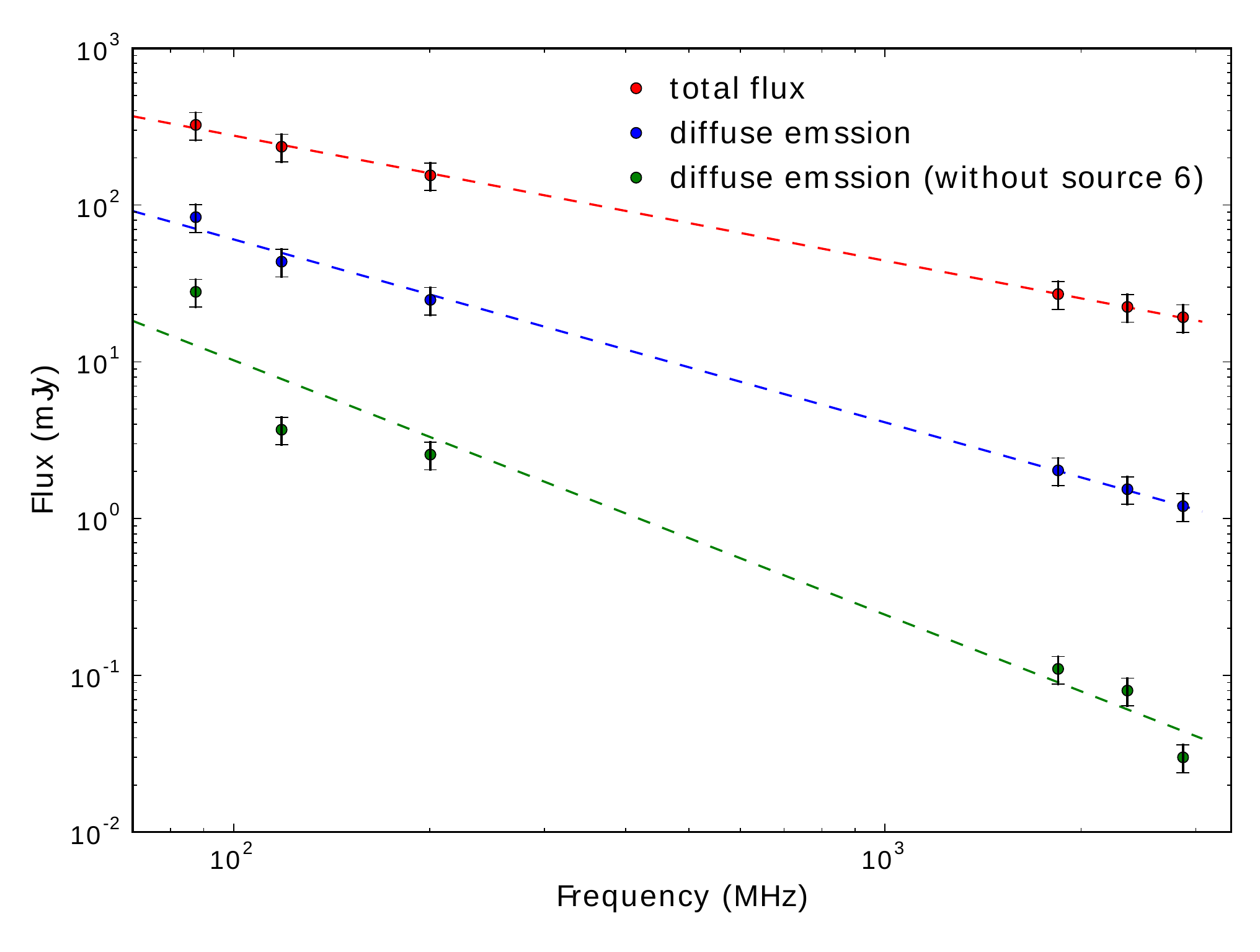} 
\end{center}
\caption{The fitted spectra of the total flux density and the flux density of the diffuse emission. The fitted spectral index of the total flux density is -0.80. The fitted spectral indices of the diffuse emission with and without Source 6 are -1.17 and -1.63, respectively.\label{fig:spectra-diffuse}}
\end{figure}

The total flux densities (discrete plus diffuse emission), the flux densities of the discrete sources and the flux densities of diffuse emission in different frequency bands from 87.5 MHz to 2.8 GHz are shown in Table \ref{tab:diffuseflux}. Figure \ref{fig:spectra-diffuse} displays the fitted spectra of the total flux density and flux density of only the diffuse emission for two scenarios; one in which all 6 discrete sources are subtracted and one in which the first 5 sources only are subtracted. Due to the poor quality of the 1332 MHz ATCA image, we do not include points from this band. Due to flux calibration issues for the 154.5 MHz sub-band MWA image as a result of increased RFI due to known satellite emission, it was not included in the calculation and hence there are only 3 MWA bands used in our analyses. The fitted spectral index of the diffuse emission when all 6 sources are removed (labeled `without Source 6') is $\alpha=-1.63_{-0.10}^{+0.10}$, which is consistent with the steep spectral indices of radio relics which fall in the range -0.9 and -2.8, with a median value of -1.3. 

\begin{deluxetable*}{cccc}
\tablecaption{The total flux density, the flux density of point sources (without (with) Source 6) and the flux density of diffuse emission of SPT J0245-5302 (with (without) Source 6) at different frequency bands from 87.5 MHz to 2.8 GHz. The fitted spectral index of the diffuse emission with source 6 included is $\alpha=-1.17_{-0.10}^{+0.10}$. The fitted spectral index of the diffuse emission without source 6 is $\alpha=-1.63_{-0.10}^{+0.10}$. Due to the flux calibration issues of the 154.5 MHz sub-band MWA image, it was not included in the calculation.\label{tab:diffuseflux}}
\tablewidth{0pt}
\tablehead{
  \colhead{Frequency } & \colhead{$S_{\rm total}$} & \colhead{$S_{\rm discrete source}$} & \colhead{$S_{\rm diffuse}$}\\
  \colhead{(MHz)} & \colhead{(mJy)} & \colhead{(mJy)} & \colhead{(mJy)}
}
\decimalcolnumbers
\startdata  
  $\rm S_{87.5}$  & 324.5$\pm$64.9 & 240.8$\pm$48.2 (296.6$\pm$59.3)& 83.7$\pm$16.7 (28.0$\pm$5.6)\\
$\rm S_{118.5}$ & 235.6$\pm$47.1 & 192.1$\pm$38.4 (231.9$\pm$46.4)& 43.5$\pm$8.7 (3.7$\pm$0.7)\\
$\rm S_{154.5}$ & - & 157.6$\pm$31.5 (187.3$\pm$37.5)& -\\
$\rm S_{200.5}$ & 154.6$\pm$30.9 & 129.8$\pm$26.0 (152.1$\pm$30.4)& 24.8$\pm$5.0 (2.6$\pm$0.5)\\
$\rm S_{1844}$  &  27.0$\pm$5.4 &  25.0$\pm$5.0 (26.9$\pm$5.4)& 2.0$\pm$0.4 (0.1$\pm$0.02)\\
$\rm S_{2356}$  &  22.4$\pm$4.5 &  20.9$\pm$4.2 (22.3$\pm$4.5)& 1.5$\pm$0.3 (0.1$\pm$0.02)\\
$\rm S_{2868}$  &  19.3$\pm$3.9 &  18.1$\pm$3.6 (19.2$\pm$3.8)& 1.2$\pm$0.2 (0.03$\pm$0.01)\\
\enddata
\end{deluxetable*}

\subsection{Discrete Source Subtraction in Images}
\begin{figure*}[ht!]
\plotone{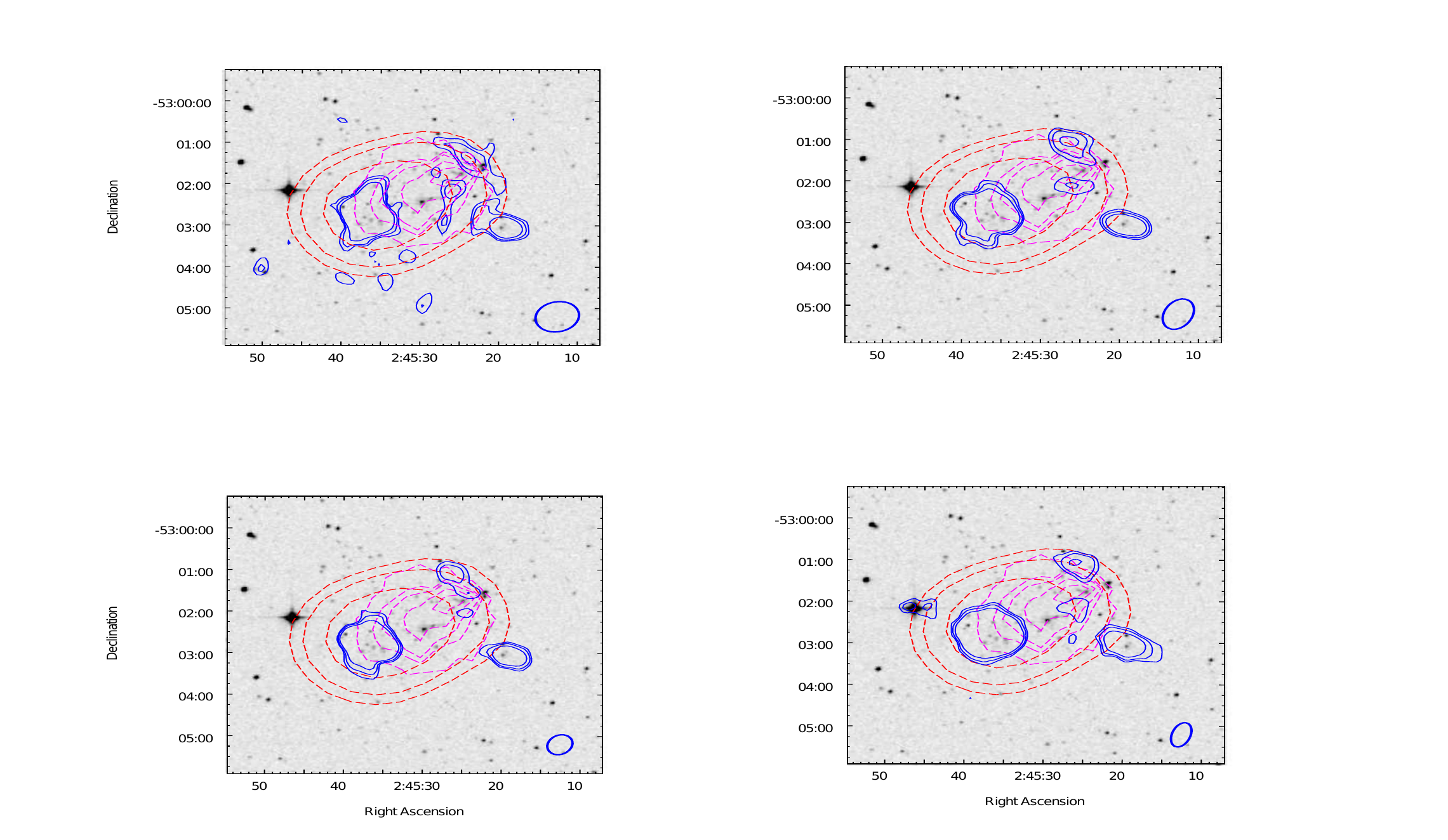}
\caption{DSS optical images of cluster SPT-CL J0235-5302 overlaid with the contours of X-ray observation (magenta), the contours of the 200 MHz MWA image (red; shown at 2$\sigma$, 3$\sigma$, and 5$\sigma$) and the contours of ATCA images after the point source subtraction (blue; shown at 2$\sigma$, 3$\sigma$, and 5$\sigma$), clockwise from top left: 1332, 1844, 2356 and, 2868MHz. The synthesised beams of ATCA images are shown as blue ellipses in the right bottom of the panels.\label{fig:subtracted}}
\end{figure*}

In order to study the morphological properties of the diffuse emission, we subtracted the four point sources in the east of the cluster from each of the ATCA sub-bands using the task UVMODEL in {\sc miriad}. The four subtracted point sources (1, 2, 3, and 5) are identified using the 6A ATCA configuration and shown in Figure \ref{fig:6A} and listed in Table \ref{tab:sources}. As previously noted, the bright source, Source 4 (02h 45m 36.64s, -53d 02m 48.23s), has complex structure (See Figure \ref{source4}) and cannot be perfectly modelled. As a result attempts to subtract this from the visibility data were not successful and it remains in the final images. Following subtraction of the discrete sources the four ATCA sub-bands were re-imaged at low resolution as described in Section \ref{sec:atca} and the resultant images are shown in Figure \ref{fig:subtracted}. 

Each panel of Figure \ref{fig:subtracted} shows the SDSS optical images overlaid with the Chandra X-ray contours (magenta) and the MWA 200 MHz contours (red) while the point source-subtracted radio emission is shown in blue contours at 1332, 1844, 2356 and, 2868MHz respectively, clockwise from the top left. Source 4 is still seen as a discrete source to the east of the cluster embedded in diffuse emission. To the west of the cluster at the tip of the X-ray emission we see a region of diffuse emission which seems to sit either side of the putative shock. We designate this as a radio relic and note that this position is exactly where shock-driven radio emission is expected to be found \citep{Skillman13}. To the southwest around the position of Source 6 is another patch of diffuse emission which at the lowest frequency is connected to the stronger patches sitting either side of the shock cone. If we consider all of the diffuse emission in the west to be a single source as in the top left panel of Figure \ref{fig:subtracted}  this suggest the relic has a size of $\sim$ 530 kpc. If only the pieces either side of the shock form the relic then the emission is approximately 320 kpc. Although Source 4 could not be subtracted, the image also clearly shows that it is embedded in larger scale diffuse emission at the other end of the X-ray emission, the diffuse component of which is exactly positioned to be a second relic in this system. The linear size of this emission is $\sim$ 370 kpc. Given the position of these regions of diffuse emission either side of the X-ray emission, following the merger axis of the cluster, along with the steep spectral index of the emission, we designate the Torpedo cluster as a double relic system at moderate redshift. 

Source 4 (SUMSS J024536$-$530244) being embedded in the diffuse emission suggested the radio relic could trace the underlying region of excess density of seed electrons remaining from older emission from the AGN. The connection between a radio galaxy and a relic has been seen several times now including in the Bullet cluster \citep{Shimwell15} and A3411 \citep{vanWeeren17} and the idea that relics are visible due to the re-acceleration of dormant seed electrons injected into the ICM by AGN has been put forward as an explanation of why these sources are not seen in every merging cluster \citep{mjh17}. Given that as with A3411 we can confirm the AGN is in the cluster itself, the connection seems likely. 

\begin{figure*}[ht!]
\begin{center}
\includegraphics[angle=0,scale=0.7]{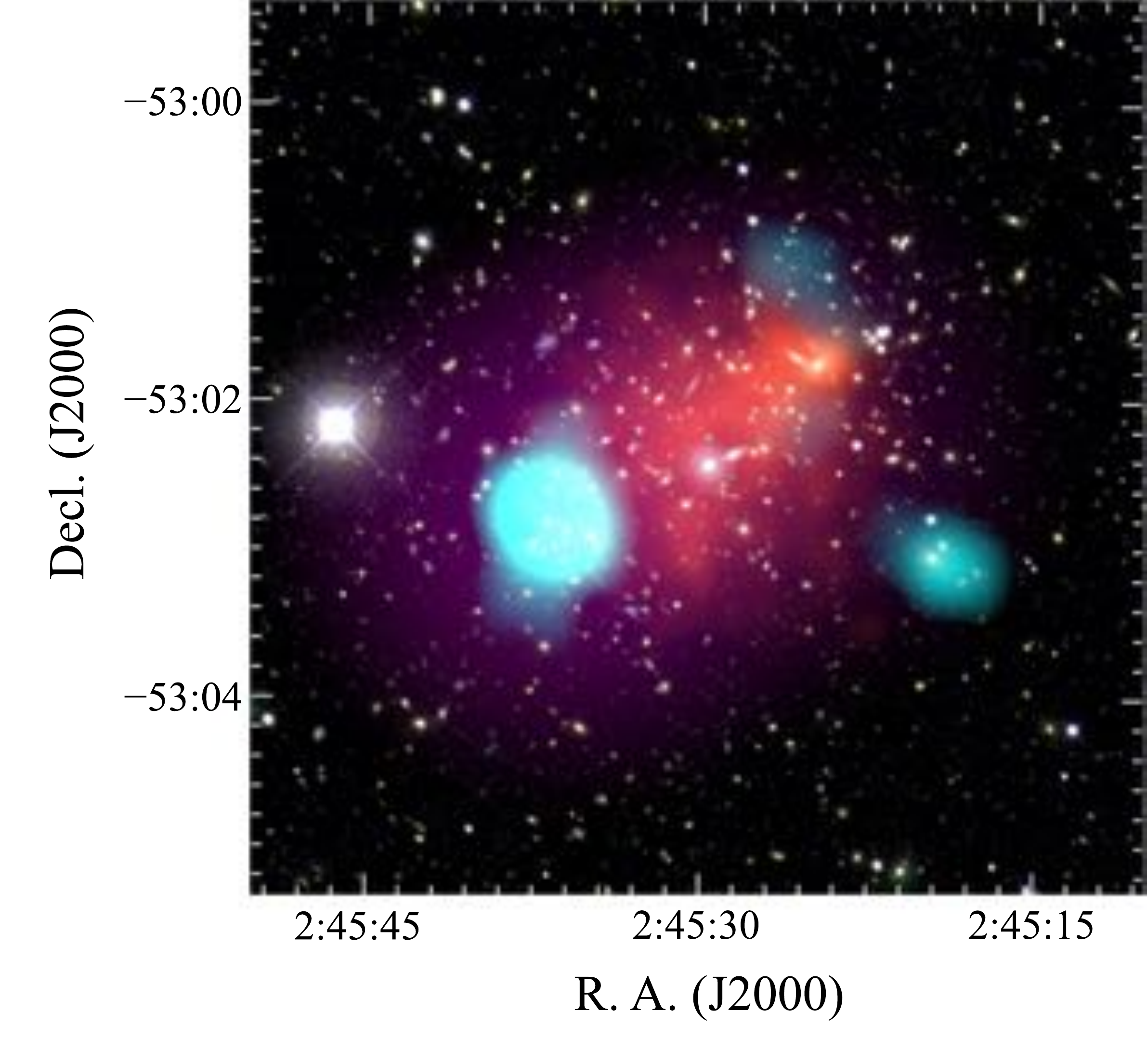} 
\end{center}
\caption{Composite image of the Torpedo cluster: the background is the multi-colour CTIO image from \protect\cite{Menanteau10}. The X-ray surface brightness from the Chandra soft-band (5.8-14.7 keV) data is shown in orange. The 200 MHz MWA observations are shown in magenta;  the beam size is 2.35 arcmin $\times$ 2.22 arcmin and data displayed cover the flux density range 32-129 mJy/beam. Finally, radio emission of the three low resolution stacked source-subtracted ATCA images is displayed in cyan, the central frequency is 2.297 GHz and the beam is 27.5" x 22.4", P.A. = 53 degrees, data shown here cover flux density values above 150 $\mu$Jy/beam.\label{fig:everything}}
\end{figure*}

Figure \ref{fig:everything} shows the combined results of the imaging undertaken here along with ancillary optical data from \cite{Menanteau10}.

\subsection{Polarization}
\begin{figure*}[ht!]
\plotone{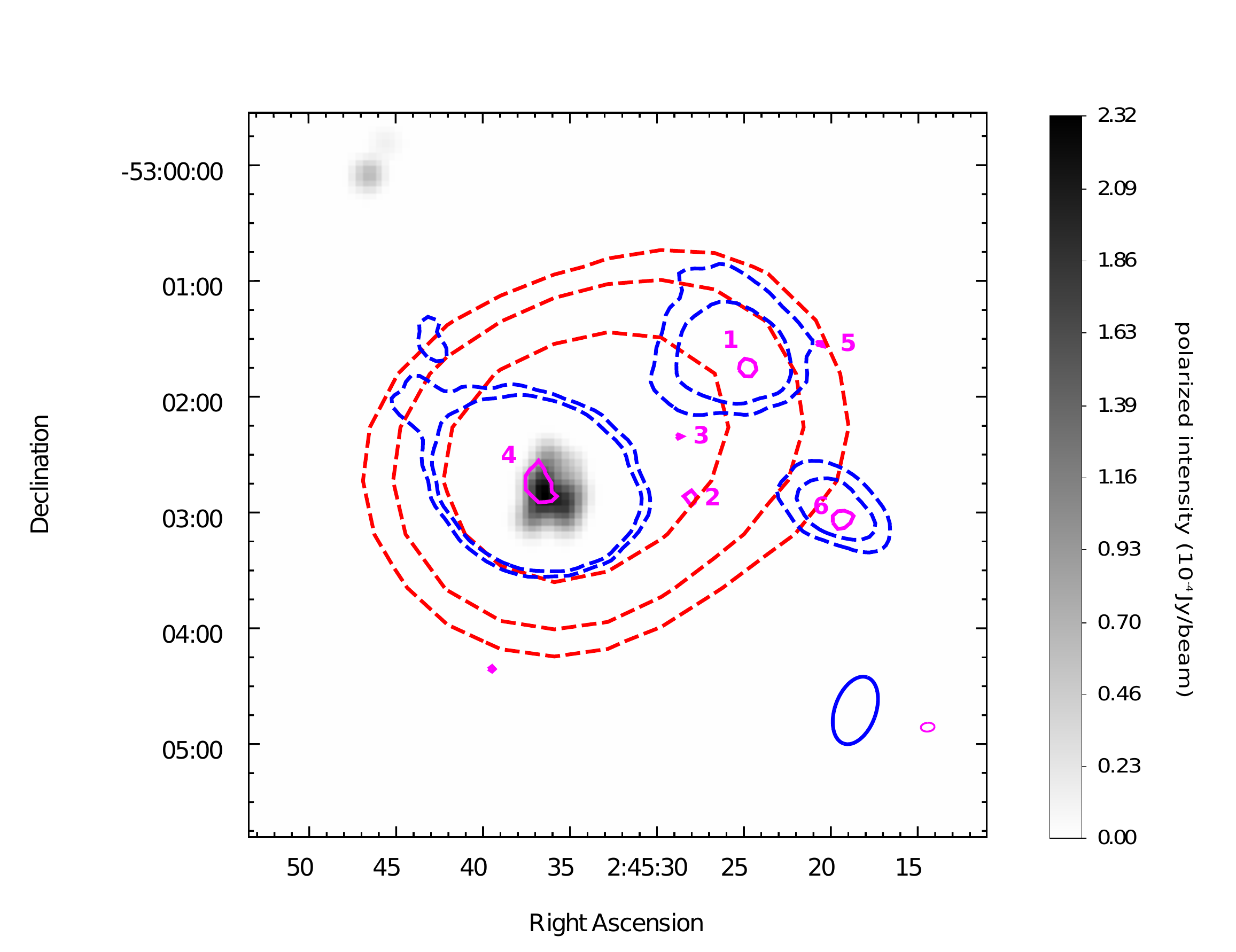}
\caption{Polarized intensity map at 2868 MHz overlaid with the contours of six point sources (magenta), the contours of 200 MHz MWA image (red, 2$\sigma$, 3$\sigma$, and 5$\sigma$) and the contours of ATCA images after the point source subtraction (blue,3$\sigma$, and 5$\sigma$).Synthesised beams of ATCA images are shown as blue ellipses in the right bottom of the panels.\label{fig:poli}}
\end{figure*}

Radio relics in galaxy clusters are strongly polarised (\citealp{mjh03,Bonafede09,Loi17}) sometimes up to 40 per cent. We would expect relics in the Torpedo cluster to also be polarised, however we are unable to measure the polarisation properties of the diffuse emission in the Torpedo cluster. The polarized intensity map is shown in Figure \ref{fig:poli}. 

No polarisation was detected in the east region of the cluster above $\sim 0.18$ mJy/beam at 2868 MHz in the eastern relic. This could be due to the depolarisation caused by the ICM, or more likely the polarised intensity is too low to be detected. Deeper observations might shed further light on both the fractional polarisation and the magnetic fields in the relic. 

Polarization of the western relic centered at (02h 45m 36s, -53d 02m 47s) was detected, with polarized intensity in the range of 292.1 $\mu$ Jy/beam to 184.2 $\mu$ Jy/beam. However, the scale of the polarised emission seems to be coincident only within the area of the known double AGN, and does not extend outwards towards the diffuse emission in which the AGN is embedded. 

\subsection{Shock Mach Number of Torpedo Cluster SPT J0245-5302}
The Mach number is a dimensionless quantity which represents the ratio of flow velocity past a boundary to the local speed of sound. The expected Mach number of the shock that produced the relic is \citep{blandford87}: 
\begin{equation}
\delta=2\frac{\rm M^2+1}{\rm M^2-1} 
\end{equation}
where $\delta$ is the average power law index of the energy spectrum of emitting electrons. $\delta$ is related to the spectral index $ \alpha$ by:
\begin{equation}
 -\alpha=(\delta-1)/2
\end{equation}
The spectral index of the elongated radio relic in the east of the cluster is  $\alpha=-1.63_{-0.10}^{+0.10}$ (without Source 6). Using this we derive a shock Mach number of $\rm M=2.04$, which is consistent with the Mach number of other detected radio relics (eg. \citealp{Hindson14}, Table 4). The asymptotic angle $\phi$ of the shock with respect to the symmetry axis satisfies:
\begin{equation}
 \rm{sin}\phi=M^{-1}
\end{equation}
For $\rm M=2.04$, the asymptotic angle $\phi \simeq 29.4$ degrees.

\section{Conclusion}
\label{sec:conclusion}

The Torpedo cluster, SPT-CL J0245-5302, at the redshift of $z = 0.300$, is a massive, merging cluster which displays a striking similarity to the Bullet cluster with soft-band X-ray data revealing it to have elongated X-ray emission with a very sharp shock cone. 

We present ATCA data from 1.1 GHz to 3.1 GHz and MWA data at frequencies between 72 MHz and 231 MHz to study the properties of SPT J0245-5302. Six discrete sources are identified in the clusters, with the the spectral indices in the range $-0.21$ and $-0.91$. One source with diffuse emission is also detected with a steep spectral index of$ -1.11_{-0.12}^{+0.06}$. The discrete sources in the west of the cluster can be modelled and subtracted from the visibility data to reveal large-scale diffuse emission in the form of a radio relic tracing the border of the X-ray shock cone. To the east of the cluster we find evidence of further diffuse emission sitting at the periphery of the X-ray emission which we believe to be a second relic. The western relic has a size of 320 - 530 kpc depending on if the south western diffuse emission is connected, while the eastern relic is 370 kpc. This eastern relic is found to have a bright cluster AGN embedded within it which we posit is the source of the seed electrons which have been re-accelerated by the passage of the merger shock. This supports the theory that mergers and the shocks they produce are a necessary but not sufficient condition to generate relics and that an underlying seed electron population deposited by AGN are the key to having a relic. The fitted average spectral index of the diffuse emission is $\alpha=-1.63_{-0.10}^{+0.10}$, which is consistent with the steep spectral indices of radio relics in the literature. 

For the detected western relic we derived the Mach number of M = 2.04 for the shock, which again is consistent with the Mach number of the previously known radio relics. The diffuse emission shows an uneven brightness distribution which is characteristic of relics, and we suspect we have not detected the full source with hints of larger emission seen in the lowest frequency ATCA imaging.  Although relics are highly polarised, we are unable to measure the polarisation properties of the diffuse emission in the cluster likely as a result of the faintness of the emission which would make the polarised intensity too faint to be measured. Further observations are warranted to study the magnetic field in the relics and around the shock. 

Taken together, the location, morphology, spectral properties of diffuse radio emission along with the clear indications of merging in the Torpedo cluster are consistent with it being a double relic system at moderate redshift. We note that few double relic systems are known at redshifts beyond 0.3 and so this is an important new detection.  

Finally, frequencies higher than 1.4 GHz are not optimal for detecting steep spectral index relics \citep{Duchesne17}, and high resolution low frequency observations would likely do a considerably better job of separating the diffuse emission from the discrete components. Unfortunately the Torpedo cluster is not accessible with the Giant Metrewave Radio Telescope and while the MWA has recently undergone a significant upgrade to double the instrument's resolution, at this redshift this is likely not to significantly improve the radio imaging of this system. Observations with SKA1 LOW are thus the most promising future prospect for obtaining a better understanding of the system.

\section*{Acknowledgements}
\addcontentsline{toc}{section}{Acknowledgements}

QZ and MJ-H are supported in this work through Marsden Fund grants. SWD is supported by an Australian Government Research Training Program scholarship administered through Curtin University. WTL is supported by the National Science Foundation of China (grant No. 11433002). This research also made use of the NASA/IPAC Extragalactic Database (NED) which is operated by the Jet Propulsion Laboratory, California Institute of Technology, under contract with the National Aeronautics and Space Administration. The Digitized Sky Surveys were produced at the Space Telescope Science Institute under U.~S.~ Government grant NAG W-2166. The images of these surveys are based on photographic data obtained using the Oschin Schmidt Telescope on Palomar Mountain and the UK Schmidt Telescope. The plates were processed into the present compressed digital form with the permission of these institutions.

\end{document}